\documentclass[preprintnumbers,two column, showpacs,amsmath]{revtex4}
\usepackage{graphicx}
\usepackage{dcolumn}
\usepackage{bm}

\begin{document}

\title{Quantum correlations in the collective spin systems}
\author{Chen Wang$^{1,2}$}
\author{Yu-Yu Zhang$^{1,3}$}
\author{Qing-Hu Chen$^{2,1,}$}\email{qhchen@zju.edu.cn}

\address{
$^{1}$Department of Physics, Zhejiang University, Hangzhou 310027,
P. R. China \\
$^{2}$Center for Statistical and Theoretical Condensed Matter
Physics, Zhejiang Normal University, Jinhua 321004, P. R. China  \\
$^{3}$Center for Modern Physics, Chongqing University£¬ Chongqing
400044, P. R.  China
 }
\date{\today}

\begin{abstract}
Quantum and classical pairwise correlations in two typical
collective spin systems (i.e., the Dicke model and the
Lipkin-Meshkov-Glick model) are discussed. These correlations in the
thermodynamical limit are  obtained analytically and in a finite-size
system  are calculated numerically. Large-size scaling  behavior for
the quantum discord itself is observed, which has never been
reported in another critical system. A logarithmic diverging behavior
for the first derivative of the quantum discord is also found in
both models, which might be universal in the second-order quantum phase transition.
It is suggested that the pronounced  maximum or minimum
of first derivative of quantum discord signifies the critical point.
Comparisons between the  quantum discord and the scaled concurrence are
performed. It is shown that the quantum discord is very small in one
phase and robust in the other phase, while the scaled concurrence shows
maximum at the critical point and decays rapidly when away from the
the critical point.

\end{abstract}


\maketitle

\section{Introduction}

Correlation, as a fundamental feature, has been extensively
investigated  in many-body physics and the quantum information
science~\cite{taylor1,nielsen1}. They indeed unravel the key
physical properties and characterize the remarkable phenomena of the
critical systems, such as quantum phase transition
(QPT)~\cite{sachdev1}. QPTs reveal the qualitative change of the
quantum systems resulting from the energy level crossings at zero
temperature.

Total correlations can be split into a classical and a quantum contribution~\cite{modi1}.
Although entanglement is only one particular kind of quantum correlation,
it has been widely applied as the characteristic measure of quantum correlations~\cite{amico1}.
Specifically, entanglement successfully identifies the critical behavior of various systems
at a QPT ~\cite{osterloh1,gu2,wu1,gu1,legeza1,julien1}.
However, the entanglement
may fail to capture the existence of the quantum correlation in some
mixed separate states, in which the entanglement is considered not a
good measure~\cite{ollivier1,oppenheim1}.
A new and alternative kind of quantum correlation based on measurement,
quantum discord (QD), is present even in separable states~\cite{ollivier1}.
From its definition, QD can be interpreted as the difference of the total correlations
(measured by the quantum mutual information) between two subsystems A and B,
before and after a local measurement performed on one of them.
The QD has been proved as a good measure of the
nonclassical correlations beyond entanglement.
Furthermore, the QD might be the source of the quantum speedup for
the mixed state quantum computation~\cite {datta1,lanyon1}.

With the capability of the QD to characterize the QPT, much attention has been
recently paid to apply the QD to quantify the critical properties of many-body systems~\cite{modi1}.
The QD or the corresponding derivatives close to the critical point generally exhibit nonanalytical
or discontinuous behaviors.
For the one-dimensional~\emph{XXZ} model at finite temperature, the entanglement fails to
pick up the critical behaviors. The discontinuity of the QD is clearly shown at the critical
points~\cite{werlang0,werlang1}, representing a benchmark for the quantum correlations in QPTs.
The critical properties of the~\emph{XXZ} model and transverse Ising model are also studied by the QD
at zero temperature, and the logarithmic scaling behaviors of the derivatives
of the QD are described~\cite{raoul1,sarandy1,amico2}.
The QD of the one-half spin~\emph{XY} chains at the critical point~\cite{maziero1} and ones
with symmetry-breaking field~\cite{tomasello1} is exhibited,
so is the QD done in the extended~\emph{XY} models~\cite{liu1,li1}.
Particularly, the bipartite QD still identifies a QPT even considering spin pairs farther than the next nearest neighboring
in Ref.~\cite{maziero1}, while the entanglement vanishes.
The exponential scaling of the QD is unraveled
to govern the ground-state factorization at the  noncritical regime in Ref.~\cite{tomasello1}.
Apart from spin systems, the QD is also used to characterize QPTs
in the correlated electron systems~\cite{allegra1} and a topological
QPT in the Castelnovo-Chamon model~\cite{yxchen}.

The Dicke model~\cite{dicke1} and  Lipkin-Meshkov-Glick (LMG)
model~\cite{LMG} are two well-known quantum collective spin models.
The Dicke model describes  the interaction of $N$ two-level atoms with
a single bosonic mode.  The LMG model  was originally introduced in
nuclear physics, but now has found applications in  other
fields~\cite{Cirac,Dusuel,Leyvraz,vidal2}.  It describes $N$  mutually
interacting spins in a transverse magnetic field.
Both models have exhibited apparent QPTs.
Recently, the collective model has been  realized experimentally in a Bose-Einstein
condensate in an optical cavity~\cite{Baumann}. A direct link
between experiments and generic models that capture QPTs has been
established~\cite{Strack}.
From the aspect of the quantum entanglement, the collective spin models have shared
the identical scaling  universality at the critical regime,
which are essentially different from spin chains~\cite{reslen1,osterloh1}.
However, the QD has not been well analyzed in these models,
except preliminarily approximate results for the QD in the
thermodynamical limit~\cite{sarandy1} and the mutual information at
finite temperatures \cite{Wilms} in the LMG model. Hence, the open question
of investigating this more general kind of the bipartite quantum correlation
naturally arises. Our paper is intended to solve the problem by studying the features
of the QD and the classical correlation for the collective spin systems.

In the present paper the QD and the classical correlation of the Dicke model
and the LMG model are investigated at zero temperature,
both in the thermodynamic limit and for finite-size systems.
The critical behavior related to
the QD and its first derivative  are analyzed. The paper is outlined
as follows. In Sec. II we review the definition of the QD and the
classical correlation, and the pairwise density matrix of the collective
spin systems is derived. In Sec. III we analyze the QD, the counterpart classical
correlation, and the entanglement in detail of the Dicke model, and
discussions are also presented. In Sec. IV, these features are
studied in the LMG model. Finally, we summarize our work in Sec. V.

\section{General formalism for quantum and classical correlations in the
collective spin models}

In classical information theory, the correlation of two subsystems
$\mathcal{A}$ and $\mathcal{B}$ can be measured by the mutual information,
which reads
\begin{eqnarray}
\mathcal{I}(\mathcal{A};\mathcal{B})=H(\mathcal{A})+H(\mathcal{B})-H(
\mathcal{A},\mathcal{B}),  \label{m:1}
\end{eqnarray}
where $H(a)=-\sum_kp_{a=k}\ln{p_{a=k}}$ ($a=\mathcal{A},
\mathcal{B}$) is the Shannon entropy, with $p_{a=k}$ the probability of the realization $k$
for the subsystem $a$. The joint Shannon entropy of $\mathcal{A}$ and
$\mathcal{B}$ is denoted as $H(\mathcal{A}
,\mathcal{B})=-\sum_{j,k}p_{\mathcal{A}=j,\mathcal{B}=k}{\ln}p_{\mathcal{A}
=j,\mathcal{B}=k}$ with $p_{\mathcal{A}=j,\mathcal{B}=k}$  the joint
probability of the subsystems $\mathcal{A}$ and $\mathcal{B}$,
realized by $ j$ and $k$. By using the Bayes rule
$p_{\mathcal{A}|\mathcal{B}=k}=p_{ \mathcal{A},\mathcal{B}=k}/
p_{\mathcal{B}=k}$, the classical mutual information can also be
rewritten into the equivalent expression
\begin{eqnarray}
~
\mathcal{J}(\mathcal{A};\mathcal{B})=H(\mathcal{A})-H(\mathcal{A}|\mathcal{
B}),  \label{m:2}
\end{eqnarray}
where $H(\mathcal{A}|\mathcal{B})=-\sum_{j,k}p_{\mathcal{A}=j,\mathcal{B}=k}
{\ln}p_{\mathcal{A}=j|\mathcal{B}=k}$ is the conditional entropy of the $%
\mathcal{A}$ and $\mathcal{B}$, and $p_{\mathcal{A}=j|\mathcal{B}=k}$ is
the corresponding conditional probability.

For quantum systems, we replace the Shannon entropy with the von Neumann entropy
in the two definitions of the mutual information.
The joint entropy is shown as $H(\mathcal{A},\mathcal{B})= -\text{Tr}_{%
\mathcal{A},\mathcal{B}}\{{\rho}_{\mathcal{A},\mathcal{B}} \ln{\rho}_{%
\mathcal{A},\mathcal{B}}\}$, where ${\rho}_{\mathcal{A},\mathcal{B}}$ is the
density matrix of the total system. Similarly, the von Neumann entropy for
the subsystem is $H(\mathcal{A})=-\text{Tr}_{\mathcal{A}}\{{\rho}_{\mathcal{%
A}} {\ln}{\rho}_{\mathcal{A}}\}$, with $\rho_{\mathcal{A}}=\text{Tr}_{%
\mathcal{B}} \{{\rho}_{\mathcal{A},\mathcal{B}}\}$. The quantum conditional
entropy quantifies the missing information of $\mathcal{A}$ after selecting
the state of $\mathcal{B}$. It can be carried out from the conditional
density matrix,
\begin{eqnarray}
{\rho}_{\mathcal{A}|{\Pi}^{\mathcal{B}}_{k}}={\Pi}^{\mathcal{B}}_{k} {\rho}_{%
\mathcal{A},\mathcal{B}}{\Pi}^{\mathcal{B}}_{k}/p_k,
\end{eqnarray}
where $\{{\Pi}^{\mathcal{B}}_{k}\}$ is a complete set of orthonormal bases
performed on the subsystem $\mathcal{B}$, with the projecting state $k$,
and the corresponding probability is $p_{k}=\text{Tr}_{\mathcal{A},\mathcal{B}}
\{{\Pi}^{\mathcal{B}}_{k}{\rho}_{\mathcal{A},\mathcal{B}}\}$.
Usually, these two mutual measurements become different, which leads
to the emergence  of the QD $\mathcal{D}({\mathcal{A}:\mathcal{B}})
=
\mathcal{I}(\mathcal{A};\mathcal{B})-\mathcal{J}(\mathcal{A};\mathcal{B})$~\cite{ollivier1,zurek1}.
In the quantum measurements, the QD between $\mathcal{A}$ and
$\mathcal{B}$ reads
\begin{eqnarray}
~ \mathcal{D}(\mathcal{A}:\mathcal{B})=\min_{\{{\Pi}^{\mathcal{B}}_{k}\}}\{
H(\mathcal{B})-H(\mathcal{A},\mathcal{B})+H(\mathcal{A}| \{{\Pi}^{\mathcal{B}%
}_{k}\})\},  \label{discord:1}
\end{eqnarray}
where the minimization is taken over the complete set of the orthogonal projectors
$\{{\Pi}^{\mathcal{B}}_{k}\}$.
Then the classical correlation is described as
\begin{eqnarray}
~ \mathcal{C}(\mathcal{A}:\mathcal{B})=\max_{\{{\Pi}^{\mathcal{B}}_{k}\}}\{H(%
\mathcal{A})-H(\mathcal{A}|\{{\Pi}^{\mathcal{B}}_{k}\})\}.
\label{classical:1}
\end{eqnarray}

Now we explain how to compute QD in collective spin systems.
The Dicke and LMG models are two typical examples. The general model can be
described as
\begin{eqnarray}
H&=&{\omega}a^{\dag}a+{\Delta}\sum_i{\sigma^{z}_{i}}+\frac{2{\lambda}}{\sqrt{N}}(a^{\dag}+a)\sum_i{\sigma^{x}_i}\nonumber\\
&&-\frac{\omega_0}{2N}\sum_{i<j}({\sigma}^{x}_{i}{\sigma}^{x}_{j}
+{\gamma}{\sigma}^{y}_{i}{\sigma}^{y}_{j}).
\end{eqnarray}
$a^{\dag}~(a)$ creates (annihilate) single photon with frequency $\omega$.
$\sigma^{k}_{i}~(k=x,y,z)$ describes the Pauli operator of $i$th ($i=1,\cdots,N$) atom, with
$N$ the number of the atoms. $\Delta$ is the energy splitting per atom.
$\lambda$ describes the atom-photon coupling strength, and $\omega_0$ is
the atom-atom interaction strength. $\gamma$ is the anisotropic factor.

Since the QD is adopted to
describe the nonlocal quantum correlation, it is necessary to derive
the pairwise density matrix. In this paper, we only study the
pairwise correlations.
For all collective spin models, including the Dicke model and the LMG model, the pairwise reduced density matrix
in the standard basis,
$\{|{\downarrow}{\downarrow}{\rangle},
|{\downarrow}{\uparrow}{\rangle},|{\uparrow}{\downarrow}{\rangle},
|{\uparrow}{\uparrow}{\rangle}\}$
(with ${\sigma_z}|{\uparrow}{\rangle}=|{%
\uparrow}{\rangle}$ and ${\sigma_z}|{\downarrow}{\rangle}=-|{\downarrow}{%
\rangle}$)~\cite{wang1},
can be derived as
\begin{equation}
~ \rho=\left(
\begin{array}{llll}
v_{+} & x_{+}^{*} & x_{+}^{*} & u^{*} \\
x_{+} & w & y & x_{-}^{*} \\
x_{+} & y & w & x_{-}^{*} \\
u & x_{-} & x_{-} & v_{-}
\end{array}
\right).  \label{rho:1}
\end{equation}
The detailed expressions for these elements are
\begin{eqnarray}
~
v_{\pm}&=&\frac{N^2-2N+4{\langle}J^{2}_{z}{\rangle}{\pm}4(N-1){\langle}
J_{z}{\rangle}}{4N(N-1)},  \label{rho:2} \\
x_{\pm}&=&\frac{(N-1){\langle}J_{+}{\rangle}{\pm}{\langle}[J_{+},J_{z}]_{+}{
\rangle}}{2N(N-1)},  \nonumber \\
w&=&\frac{N^2-4{\langle}J^2_{z}{\rangle}}{4N(N-1)},
y=\frac{{\langle}
J^2_x+J^2_y{\rangle}-N/2}{N(N-1)},  \nonumber \\
u&=&\frac{{\langle}J^2_+{\rangle}}{N(N-1)},  \nonumber
\end{eqnarray}
where $[A,B]_{+}=AB+BA$. $w=y$, for
$\sum_{\alpha=x,y,z}J^2_{\alpha}=J^2= \frac{N}{2}(\frac{N}{2}+1)$.
In particular, since the Dicke and LMG models have symmetric ground states
with parity conservation, we find $x_\pm = 0$~\cite{wang1,julien1}.
Hence the pairwise reduced density matrix in $X$ form is shown as
\begin{equation}
~ \rho=\left(
\begin{array}{llll}
v_{+} &0 & 0 & u^{*} \\
0 & w & y & 0 \\
0 & y & w & 0 \\
u & 0 & 0 & v_{-}
\end{array}
\right).  \label{rho:2}
\end{equation}
For the two-qubit states in $X$ form, the QD may be derived analytically, according
to Ref.~\cite{luo1}.

From the definition in Eq.~(\ref{discord:1}), the QD $\mathcal{D }$
and the classical correlation $\mathcal{C}$ can be obtained within
the reduced subsystem von Neumann entropy ${H}_{\mathcal{A}}$, the
joint entropy ${H}_{\mathcal{A},\mathcal{B}}$, and the conditional
entropy ${H}_{ \mathcal{A}|{\Pi}^{\mathcal{B}}_{k}}$, which are
analyzed in detail in Appendix A.

\section{Quantum Correlation of the Dicke Model}

We study the QD in the Dicke model in this section. The Dicke
Hamiltonian can be written in terms of the collective momentum
~\cite{Emary,chen1},
\begin{eqnarray}
~ H_{\text{Dicke}}={\omega}a^{\dag}a+{\Delta}J_z+\frac{2{\lambda}}{\sqrt{N}}%
(a^{\dag}+a)J_x,  \label{dicke:1}
\end{eqnarray}
where $a^{\dag}$ and $a$ are the bosonic annihilation and creation
operators of the single-mode cavity, $\Delta$ and $\omega $ are the
transition frequency of the qubit and the frequency of the single
bosonic mode, and $\lambda $ is the coupling constant. $J_x$ and $J_z$
are the collective spin operators. It is well known that this model
undergoes a second-order QPT from the normal phase to the
superradiant phase, separated by the critical point
$\lambda_c=\sqrt{{\omega}{\Delta}}/2$.

We first apply the Holstein-Primakoff transformation to change the
collective angular operators to the boson operators $b~(b^{\dag})$ by $J_{+}=b^{\dag}\sqrt{%
N-b^{\dag}b}$, $J_{-}=\sqrt{N-b^{\dag}b}b$, and $J_{z}=b^{\dag}b-N/2$, where
$[b,b^{\dag}]=1$ \cite{Emary}. Then the displacements of the boson operators are
introduced to depict the behaviors of the superradiation phase as $%
c^{\dag}=a^{\dag}+\sqrt{N}{\alpha}$ and
$d^{\dag}=b^{\dag}-\sqrt{N}{\beta}$. By using large $N$ expansions
of $H_{\text{Dicke}}$ with respect to the new boson operators
$c^{\dag}$ and $d^{\dag}$ up to the $1/N$, we obtain the energy
expectation,
\begin{eqnarray}
\frac{E_{G}(\alpha,\beta)}{N}={\omega}{\alpha}^2-4{\lambda}{\alpha}{\beta}%
\sqrt{1-{\beta}^2}+{\Delta}({\beta}^2-1/2).
\end{eqnarray}
Minimizing the energy gives
\begin{eqnarray}
{\omega}{\alpha}-2{\lambda}{\beta}\sqrt{1-{\beta}^2}&=&0, \nonumber\\
2{\alpha}{\lambda}\sqrt{1-{\beta}^2}-\frac{2{\alpha}{\lambda}{\beta}^2}{%
\sqrt{1-{\beta}^2}}-{\beta}{\Delta}&=&0,
\end{eqnarray}
then we have
\begin{eqnarray}
\beta^2&=&\max\{0,\frac{1}{2}(1-{\lambda}^2_c/{\lambda}^2)\},\nonumber \\
\alpha&=&\frac{2\lambda}{\omega}{\beta}\sqrt{1-\beta^2},
\end{eqnarray}
where the transition point $\lambda_c=\sqrt{{\omega}{\Delta}}/2$.
Next we can derive the matrix elements of the pairwise reduced
density in Eq.~(\ref {rho:2}) up to $O(1)$,
\begin{eqnarray}  \label{parameter:1}
v_{+}&=&\beta^4,~v_{-}=(1-{\beta}^2)^2, \nonumber\\
w&=&y={\beta}^2(1-{\beta}^2),~u={\beta}^2(1-{\beta}^2).
\end{eqnarray}
The von Neumann entropy of the subsystem $
\mathcal{A}$ and $\mathcal{B}$ are thus given by
\begin{eqnarray}
H(\mathcal{A}(\mathcal{B}))&=&-{\beta}^2{\ln}{{\beta}^2}-(1-\beta^2)\ln(1-%
\beta^2),~  \label{ha:1} \\
H(\mathcal{A},\mathcal{B})&=&
-[(\beta^4+(1-\beta^2)^2)]{\ln}[\beta^4+(1-\beta^2)^2]\label{hab:1}\nonumber\\
&&-[2{\beta}^2(1-{\beta}^2)]\ln[2{\beta}^2(1-{\beta}^2)].
\end{eqnarray}
Note that the quantum mutual information can be obtained by $\mathcal{I}(\rho_{
\mathcal{A},\mathcal{B}})=H(\mathcal{A})+H(\mathcal{B})-H(\mathcal{A},%
\mathcal{B})$.
Then following the detail analysis in Appendix B, the minimum of the
von Neumann conditional entropy can be obtained by
\begin{eqnarray}  \label{hc:2}
H(\mathcal{A}|\Pi^{\mathcal{B}})&=&\ln(2)-\frac{1}{2}[(1+M)\ln(1+M)~
\nonumber  \label{hc:1} \\
&&+(1-M)\ln(1-M)],
\end{eqnarray}
with
$M=[(2{\beta}^2-1)^2+16{\beta}^4(1-{\beta}^2)^2]^{1/2}$.

Combining Eqs.~(\ref{ha:1})-(\ref{hc:2}), we
derive the QD as
\begin{eqnarray}~\label{qd:1}
\mathcal{D}&=&-{\beta^2}\ln{\beta^2}-(1-{\beta^2})\ln({1-\beta^2}) \\
&&+[2{\beta^2}(1-{\beta^2})]\ln[2{\beta^2}(1-{\beta^2})]  \nonumber \\
&&+[{\beta^4}+(1-{\beta^2})^2]\ln[{\beta^4}+(1-{\beta^2})^2]+\ln2
\nonumber \\
&&-\frac{1}{2}[(1+M)\ln(1+M)+(1-M)\ln(1-M)].  \nonumber
\end{eqnarray}
And the classical correlation is
\begin{eqnarray}~\label{classical:1}
\mathcal{C}&=&-{\beta^2}\ln{\beta^2}-(1-{\beta^2})\ln({1-\beta^2})-\ln2 \\
&&+\frac{1}{2}[(1+M)\ln(1+M)+(1-M)\ln(1-M)].  \nonumber
\end{eqnarray}

We next investigate the  quantum and classical correlation in the
finite-size Dicke model. Two of the present authors and
collaborators have proposed a numerically exact technique to the
Dicke model up to a very huge size by using the basis of
extended coherent states~\cite{chen1}. This effective approach has been confirmed
recently by comparing with the results in terms of the basis of the Fock
states~\cite{Hirsch}. It was demonstrated that it is
very difficult to obtain convergent results for a large number of
atoms based on the usual basis of the Fock states~\cite{Hirsch,lambert1}.

In the numerically exact approach~\cite{chen1}, the wave function can be
expressed in terms of the basis   $\{\left| \varphi _n\right\rangle
_b\bigotimes \left| j,n\right\rangle \}$ where $\left| j,n\right\rangle $ $\
$ is the Dicke state with $j=N/2$ and $\left| \varphi _n\right\rangle _b$ is
the bosonic extended coherent state
\begin{equation}
\left| \varphi _n\right\rangle _b=\sum_{k=0}^{N_{tr}}c_{n,k}\frac
1{\sqrt{k!} }(a^{\dag}+g_n)^ke^{-g_na^{\dag}-g_n^2/2}\left| 0\right\rangle
_a, \label{wavefunction}
\end{equation}
where $g_n=2\lambda n/\omega \sqrt{N}$, $N_{tr}$ is the truncated
bosonic number in the space of the new operator $A_n=a+g_n$,
$|0{\rangle}_a$ is the vacuum as $a|0{\rangle}_a=0$,
and the coefficient $ c_{n,k}$ can be determined through the Lanczos
diagonalization with errors less than $10^{-6}$.
Then, we can derive the elements of pairwise density matrix in Eq.~(\ref
{rho:1}).
Without loss of generality, we mainly focus on
the resonant case $\Delta =\omega $ in the following.

\begin{figure}[tbp]
\includegraphics[scale=0.55]{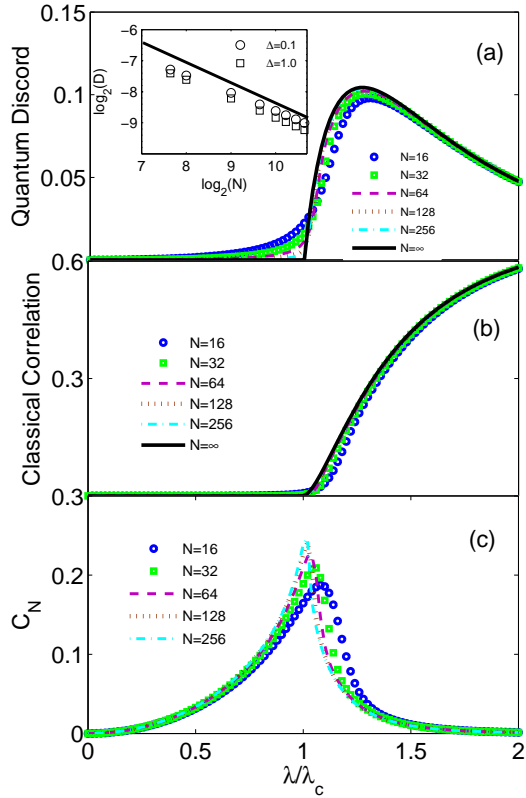}
\caption{(Color online) (a) QD, (b) classical correlation, and (c)
scaled concurrence  as a function of the coupling constant in   the Dicke
model with different sizes  for $\Delta=1.0$ and $\omega=1.0$. Inset
in (a) is the finite-size scaling of the QD at the critical point
$\lambda_c$, and the solid line scales as $N^{-2/3}$. }
\label{figure1}
\end{figure}

It is well known that the second-order QPT occurs in the
thermodynamic limit. The quantum correlation is deeply related to
QPT.
The reminiscences of the QPT for a finite (large) system size should be very interesting.
Therefore, we study the size dependence of the
QD and classical correlation in the Dicke model. In
Figs.~\ref{figure1}(a) and~\ref{figure1}(b), we display  the quantum correlation
and classical correlation as a function of the atom-cavity coupling
constant for several system sizes. The results in the thermodynamic
limit are also listed.

It is shown that both the quantum correlation and classical
correlation are quite small in the normal phase
($\lambda<\lambda_c$). They tend to zero with the increase of the
atomic number, agreeing well with the results in the thermodynamic
limit, which is exactly zero. In the thermodynamic limit, there is
no excitation of the system in the normal phase, so a correlation
between two arbitrary atoms, which can  only be mediated by the
photons, should be absent. While in the superradiant phase
($\lambda>\lambda_c$), the QD shows nonmonotonous behavior as the
coupling strength  increases in Fig.~\ref{figure1}(a). The maximum
of the QD is away from the critical point, even in the thermodynamic
limit, locating at $1.28\lambda_c$. This suggests that the nonclassical
atom-atom correlation becomes strongest at the mediate coupling regime
above the critical point. It is different from the one-dimensional~\emph{XXZ}
model~\cite{raoul1}, where the maximum of the QD is
right at the critical point.

It is very interesting to observe a power-law scaling
$\mathcal{D}|_{\lambda=\lambda_c}{\propto}N^{-{\mu}}$ for different
detunings, as shown in the inset of Fig.~\ref{figure1}(a).  The
asymptotic slop in the log-log scale in the large-size atomic
systems for different detunings   suggests  the same exponent
$\mu=2/3$. To the best of our knowledge,  such a finite-size scaling
for the QD itself has never been reported in other systems at critical regime.
In the recent study on one dimensional \emph{XY} model with symmetry-breaking
longitudinal field in Ref.~\cite{tomasello1}, an exponential scaling is observed near the
factorization points, which provides a nontrivial scaling feature of the QD even in
the noncritical regimes. This is significantly different from the results
in the Dicke model, which holds at the critical point. Nevertheless, the ground state of this \emph{XY} system
at the factorization points is separable, resulting in the disappearance of the QD. This is similar
to what happens in the Dicke model at the normal phase, where the atoms part of the ground state is
dominated by the separable form ${\Pi^N_{k=1}}|{\downarrow}{\rangle}_{k}$,
with $\sigma^{z}_{k}|{\downarrow}{\rangle}_{k}=-|{\downarrow}{\rangle}_{k}$.

As the coupling strength increases further, the QD
becomes smaller. In the strong coupling limit, the QD
finally decreases to zero, which means the nonlocal quantum
correlation of the system disappears.
In this limit ($\lambda{\gg}\lambda_c$), the ground
state can be described by
$|\psi^{\pm}_G{\rangle}=\frac{1}{\sqrt{2}}
[(\Pi^{N}_{k=1}|{e_x}{\rangle}_{k})|0{\rangle}_{A_{-N/2}}
{\pm}(\Pi^{N}_{k=1}|{g_x}{\rangle}_{k})|0{\rangle}_{A_{N/2}}]$
under $C_2$ symmetry~\cite{Hirsch},
where ${\sigma^x_k}|e_x{\rangle}_k=|e_x{\rangle}_k$
and ${\sigma^x_k}|g_x{\rangle}_k=-|g_x{\rangle}_k$.
$|0{\rangle}_{A_{-N/2}}$ and $|0{\rangle}_{A_{N/2}}$ are the vacuum for
the coherent state modes $A_{-N/2}|0{\rangle}_{A_{-N/2}}=0$
and $A_{N/2}|0{\rangle}_{A_{N/2}}=0$, respectively.
Hence, the reduced bipartite density matrix of the arbitrary two atoms
corresponds to complete einselection~\cite{ollivier1}, where the QD appears zero.
On the other hand,
it can also be seen in Eq.~(\ref{qd:1}) that $\beta^2=1/2$ as
$\lambda{\rightarrow}\infty$, so $\mathcal{D}{\rightarrow}0$.

The classical correlation in the superradiant phase  increases
monotonously  with the coupling strength, similar to those in the
different spin chains~\cite{raoul1,liu1}.  In the thermodynamical
limit, $\beta^2{\rightarrow}1/2$ in the deep coupling regime, so
that $H(\mathcal{A})=H(\mathcal{B})=\ln2$, and
$H(\mathcal{A},\mathcal{B})=\ln2$. For the conditional von Neumann
entropy, $H(\mathcal{A}|\Pi^\mathcal{B})=0$. Hence, the classical
correlation saturates at  $\log2$ in the strong coupling limit.

It is well known that the concurrence and the QD are both good
measures to investigate the nonclassical correlations. Hence,
comparisons between these two quantities are highly desirable. The
scaled concurrence has been calculated previously
by $C_{N}=1-4{\langle}J^2_y{\rangle}/N$~\cite{lambert1}.
The scaled concurrence in the very large system size has been calculated
later by two present authors and collaborators~\cite{chen1}. We also
collect  the scaled concurrence as a function of coupling constant for
different size in   Fig.~\ref{figure1} (c) for convenience.

Interestingly, we find that QD and scaled concurrence show essentially
different behaviors in the whole coupling regime. First,
it is shown that the QD decreases with the atomic number in
the normal phase and vanishes exactly in the thermodynamic limit.
While the scaled concurrence increases with the atomic number in the normal phase,
it is always finite  in the thermodynamic limit.
Since the scaled concurrence is defined by $C_N=(N-1)C_n$, where
$C_n$ is the normal concurrence, it is given by a two-point
entanglement measure times the system size (the atomic number).
On the contrary, the QD is a measure of two-point nonclassical
correlations and it is not multiplied by the atomic number.
This clearly explains their different behaviors with respect to the
increase of the system size.
The normal concurrence $C_n$ is vanishing in the thermodynamic limit,
implying that entanglement between two arbitrary atoms should be absent
in this limit. This feature is consistent with the disappearance of the QD.
Second, the scaled concurrence reaches its maximum
at the critical point where the QD is still
very small. Third, in the superradiant phase,  the QD shows a broad
maximum and decays slowly. The scaled concurrence decreases
monotonically and rapidly with the coupling constant in the strong
coupling regime. For example, for ${\lambda}/{\lambda_c}=2.0$, the
scaled concurrence becomes negligible, and the value of the QD still remains
about half of the maximum, demonstrating the persistence of the QD
in regimes where entanglement is absent, a feature already observed in the extended
Hubbard model~\cite{allegra1}.
Therefore we conclude that the scaled concurrence is not sufficient to
quantify the nonclassical correlation.
In this sense, one can say that QD is a pretty good measure of the
quantum correlation independent of entanglement.

\begin{figure}[tbp]
\includegraphics[scale=0.45]{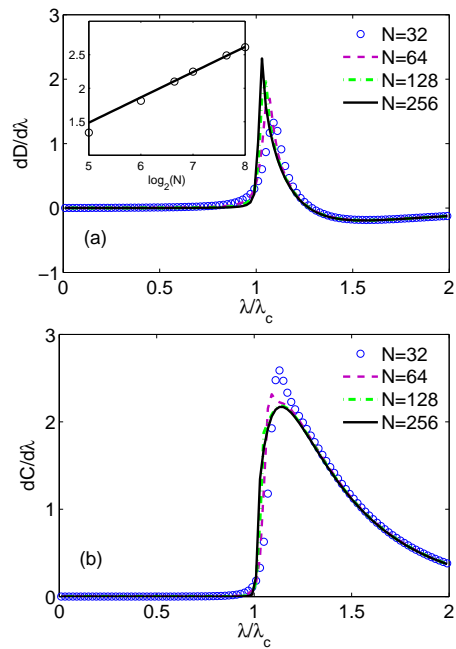}
\caption{(Color online) First  derivative of the QD (a) and the
classical correlation (b) as a function of the coupling constant in
the Dicke model with different system sizes for $\Delta=1.0$ and
$\omega=1.0$. Inset in (a) shows the finite-size scaling of the
maximum of the $d\mathcal{D}/d\lambda$. } \label{figure2}
\end{figure}

To  quantify the QPT of the Dicke model, the first derivative of the
QD and the classical correlation as a function of the coupling
constant are presented in Fig.~\ref{figure2}. The cusplike peak of
$\frac{d\mathcal{D}}{ d\lambda}$ curves emerges with an increase of the
atomic number, indicating a nonanalytical behavior.
This feature confirms a general paradigm developed by Wu \emph{et al.}~\cite{wu1}.
The peak position tends to the  critical point with increasing $N$. It would
provide a good tool to detect the critical point by a finite-size
study. In the critical spin chains, the first derivative of the QD  also
shows its peak right at the critical point, which is also
nonanalytical and discontinuous~\cite{sarandy1,maziero1,tomasello1,liu1,li1}.

Interestingly, the logarithmic scaling of the maximum of the
$d\mathcal{D}/d\lambda$ is also observed in the large size regime,
which can be fitted well by
$(\frac{d\mathcal{D}}{d\lambda})_{\textrm{max}}=0.377\ln_2(N)-0.401$,
demonstrated in the inset of Fig.~\ref{figure2}(a). This logarithmic
scaling has been also reported in Refs.~\cite{li1,sarandy1} for \emph{XY}
spin chains and transverse field Ising chains, implying the universal features
of the derivatives of the QD to exhibit the novel behaviors of the critical systems.
The first derivative of the classical correlation $\frac{d\mathcal{C}}{d\lambda}$ displays
a different behavior. It becomes more rounded around the maximum
with the increase of the size.

\section{Quantum Correlation of the LMG Model}
We turn to the other quantum collective spin model, the LMG model. Its
Hamiltonian reads~\cite{LMG,Dusuel,Leyvraz}
\begin{eqnarray}
H_{\text{LMG}}=-\frac{1}{2N}\sum_{i<j}({\sigma}^{x}_{i}{\sigma}^{x}_{j}+{
\gamma}
{\sigma}^{y}_{i}{\sigma}^{y}_{j})-\frac{\lambda}{2}\sum_{i}{\sigma}
^{z}_{i}.
\end{eqnarray}
where $\sigma_i  (i=x,y,z)$ are Pauli spin-$1/2 $ operators,
$\lambda$ is the magnetic field, and $\gamma $ is the anisotropic parameter. In
the framework of the collective spin operators
$J_k=\sum_{i}{\sigma^{k}_{i}} /2$ with $k=x,y,z$, the model can be
rewritten as
\begin{eqnarray}
H_{\text{LMG}}&=&-{\lambda}J_z-\frac{1}{N}[J^2_{x}+{\gamma}%
J^2_{y}-N(1+\gamma)/4].
\end{eqnarray}
We focus on the case of  $\gamma=0$ and $\lambda>0$ where the
second-order QPT can occur.

We also first study the quantum and classical correlation in the
thermodynamic limit.   Based on the Holstein-Primakoff
transformation $J_z=b^{\dag}b-N/2, J_{+}=b^{\dag}\sqrt{N-b^{\dag}b},
J_{-}=\sqrt{N-b^{\dag}b}b$, we displace the boson operator to the
form of $c^{\dag}=b^{\dag}+\alpha$. By large $N$ expansion of
$H_{\text{LMG}}$, the ground-state energy per spin is shown as
\begin{eqnarray}
E_G/N=-[(1-\alpha^2)\alpha^2+\lambda(\alpha^2-1/2)].
\end{eqnarray}
By minimizing $E_G$, we find
\begin{eqnarray}
\alpha^2=\min\{1,\frac{1+\lambda}{2}\}.
\end{eqnarray}
Consequently, we have ${\langle}J_z{\rangle}/N=(\alpha^2-1/2), {\langle}%
J^2_z{\rangle}/N^2=(\alpha^2-1/2)^2$, and ${\langle}J^2_+{\rangle}%
/N^2=\alpha^2(1-\alpha^2)$. Then the elements of the pairwise
density matrix in Eq.~(\ref{rho:2}) up to $O(1)$ are derived by
\begin{eqnarray}
v_+&=&(\frac{1+\lambda}{2})^2,~v_-=(\frac{1-\lambda}{2})^2,\\
w&=&y=(\frac{1+\lambda}{2})(\frac{1-\lambda}{2}),~u=(\frac{1+\lambda}{2})(\frac{1-\lambda}{2}).  \nonumber
\end{eqnarray}


By a suitable reparametrization we cast the matrix elements
in the same form as those of the Dicke model in Eq.~(\ref {parameter:1})
\begin{eqnarray}
\beta^2&=&\frac{1-\lambda}{2},~1-\beta^2=\frac{1+\lambda}{2}. \label{mapping:1}
\end{eqnarray}
Therefore the elements can be expressed as
\begin{eqnarray}
~ v_+&=&(1-\beta^2)^2,~v_-=\beta^4,  \label{mapping:2} \\
w&=&y=\beta^2(1-\beta^2),~u=\beta^2(1-\beta^2).  \nonumber
\end{eqnarray}

The influence of the anisotropic parameter $\gamma$ on the quantum
correlation is numerically checked, and we find that the
QD is almost not altered with $\gamma$.  Hence we mainly focus on
$\gamma=0$ in the following.

\begin{figure}[tbp]
\includegraphics[scale=0.55]{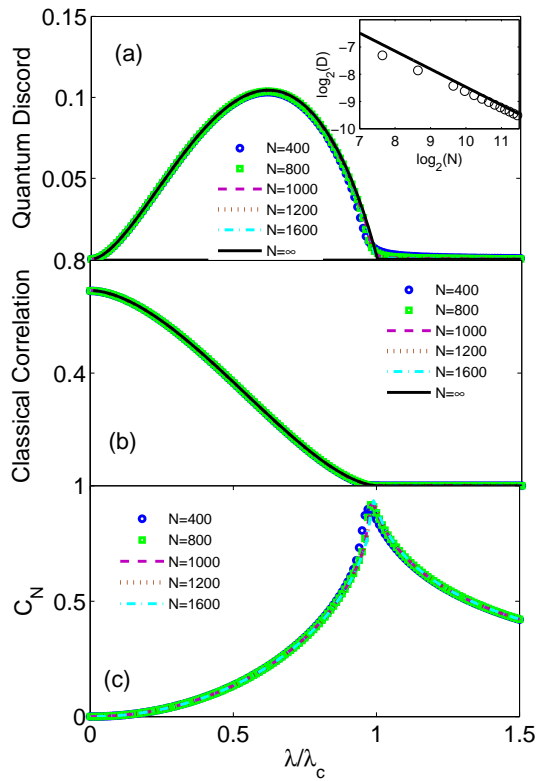}
\caption{(Color online) (a) QD, (b)  classical correlation,  and (c)
scaled concurrence as a function of the field in the LMG model with
different system sizes for $\gamma=0$. Inset in (a) is the finite-size
scaling of the QD at the critical point $\lambda_c$, and the
solid line scales as $N^{-2/3}$.  } \label{figure3}
\end{figure}

The finite-size LMG model can be studied by the exact
diagonalization on the basis of the collective spin
operators~\cite{Dusuel}. For the ground state,
convergent results can be obtained for very a large system size.
The size dependence
of the QD and the classical correlation are given in Fig.~\ref{figure3}.
The QD shows the nonmonotonous behavior in the symmetry-broken
phase ($ \lambda<\lambda_c$). Starting at $0$, the QD then shows the
single broad maximum at $\lambda=0.62\lambda_c$, far from the
critical point. The QD is very small in the symmetry phase
($\lambda>\lambda_c$). The classical correlation decreases
monotonously in the symmetry-broken phase, approaching to zero at
the critical point. If the magnetic field $\lambda=0$, the classical
correlation is easily obtained as  $\ln2$. In the symmetry
phase, the classical correlation is rather small for a finite-size
system and  becomes zero in the   thermodynamic limit.

Similar to the Dicke model, the QD of the LMG model also exhibits
the power-law scaling behavior  as
$\mathcal{D}_{\lambda=\lambda_c}{\propto}N^{-\mu}$ at the critical
point, as shown in the inset of  Fig.~\ref{figure3}(a).  The scaling
exponent $\mu$ in the large $N$ regime is very close to $2/3$, the
same as that obtained above in the Dicke model, providing a new
piece of evidence of the same universality class of these two
models.

The QD in the LMG model in the thermodynamic limit has been
preliminarily studied in Ref.~\cite{sarandy1}. By mapping the LMG model
to the two-band fermion model exactly, the quantum correlation and
the classical correlation of the fermions were studied
alternatively, based on the corresponding density matrix. It was
found that both the QD and the classical correlation decrease
monotonously in the symmetry-broken phase with the magnetic field,
and keep zero in the symmetry-broken phase. In the present paper,
the pairwise density matrix in the LMG model is evaluated directly for
qubits (i.e., spins) and it is related, but not identical to that
of Ref.~\cite{sarandy1}. The different definitions may account for
the different values of QD obtained in the two papers.
Our results for the classical correlation are consistent qualitatively
with those in Ref.~\cite{sarandy1}.

The comparisons of the QD and the scaled concurrence in the LMG
model are also made.  The scaled concurrence for large system size
has been calculated by Dusuel \emph{et al.}~\cite{Dusuel1},  which
was shown in their Fig. 9. For completeness,
we recalculate the concurrence, also including much larger system sizes,
and list them in
Fig.~\ref{figure3}(c). At the critical point, the concurrence shows
the maximum, while the QD becomes $0$. In the symmetry phase
($\lambda>\lambda_c$), the QD is very small, implying no quantum
correlation. The concurrence remains finite  and decreases
gradually and monotonously with the field $\lambda$.

\begin{figure}[tbp]
\includegraphics[scale=0.40]{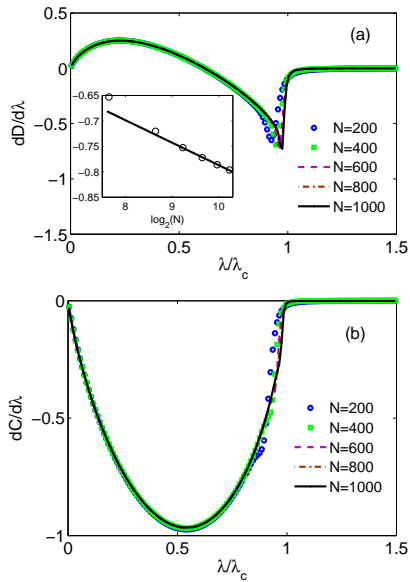}
\caption{(Color online) First-order derivative of the QD (a)  and
the classical correlation (b) as a function of the field in the LMG
model with different system sizes for $\gamma=0$. Inset in (a)
exhibits the finite-size scaling of the minimum of the
$d\mathcal{D}/d\lambda$. } \label{figure4}
\end{figure}

In order to study the critical behaviors of the LMG model, we
display the first  derivative of the QD and classical correlation in
Fig.~\ref{figure4}. $\frac{d\mathcal{D}}{d\lambda}$ has a pronounced
minimum  around  the   critical point. Note that the minimum
position approaches the critical point, providing a convincing
method to locate the critical point.
This behavior strongly mirrors the sudden transition
from the symmetry-broken phase to the symmetric phase.
The finite-size scaling of the minimum of the
$d{\mathcal{D}/d\lambda}$ is also performed in the inset of
Fig.~\ref{figure3}. The  logarithmic divergence fitted by
$(\frac{d\mathcal{D}}{d\lambda})_{\textrm{min}}=-0.044\ln_2(N)-0.346$
is also well exhibited in the large size regime, similar to that in
the Dicke model. It should be pointed out that the coefficients of
$\ln_2(N)$ in the two models are different. The first derivative of the
classical correlation shows
the broad valley around $0.55\lambda_c$, and becomes $0$ as the
field parameter $\lambda\ge\lambda_c$.
Combined with the
similar scaling behavior reported in spin chains~\cite{li1,sarandy1},
such a logarithmic divergence of the first derivative of the QD may
be universal in critical systems; further extensive confirmations in
other systems are, however, needed.

\section{Summary}

The QD and the classical correlation are  investigated to
characterize QPTs in the Dicke model and the LMG model. These
correlations in the thermodynamic limit have been derived
analytically and in the finite-size system obtained numerically
up to very a large system size. Perfect power-law scaling behavior at
the critical point is observed in both models. Such a scaling
behavior has not been reported in other critical systems,
except the exponential scaling at noncritical regimes in
the \emph{XY} chain with symmetry-breaking field~\cite{tomasello1}.
The same scaling exponents provide new evidence of the
same universality of two models. The position of the pronounced
maximum or minimum of the first derivative of the QD approaches to
the critical point with the increase of the system size, unlike some
other critical systems.
We have convincing evidence of a logarithmic diverging behavior
for the first derivative of the QD in the two models.
It is suggested that this  logarithmic diverging behavior
at the critical point might be universal in the second QPTs. The
coefficient of the logarithmic term is, however, model dependent.

We also find that the QD and the scaled concurrence show essentially
different behaviors in both models.
In the symmetry phase, the QD vanishes in the thermodynamic limit, while
the scaled concurrence remains finite. In the symmetry-breaking phase,
the QD shows the broad maximum and becomes zero in the strong
atom-cavity (atom-atom) coupling limit,
whereas the scaled concurrence decreases monotonically.
This explicitly shows the robustness of the QD in the
symmetry-breaking phase, compared to the scaled concurrence.
Recently, we note that the multipartite measure (global quantum
discord) emerges as a powerful tool for quantum
correlation~\cite{rulli1}. Hence, it may be interesting to apply the
global quantum discord to further exploit the novel behaviors in
collective spin systems in future research.

\section{Acknowledgement}
This work was supported by National Natural Science Foundation of
China

\appendix

\section{Derivation of the quantum correlation}

Considering the pairwise atom reduced density matrix in
Eq.~(\ref{rho:2}), the reduced density matrix ${\rho}_{\mathcal{A}}$
is obtained by
\begin{eqnarray}
{\rho}_{\mathcal{A}}&=&\text{Tr}_{\mathcal{B}}\{{\rho}_{\mathcal{A},\mathcal{
B}}\}  \nonumber \\
&=&(v_{+}+w)|\downarrow{\rangle}_{\mathcal{A}}{\langle}\downarrow|+(v_{-}+w)|\uparrow{\rangle}_{
\mathcal{A}}{\langle}\uparrow|.  \nonumber
\end{eqnarray}
The Von Neumann entropy of the reduced system $\mathcal{A}$ is shown as
\begin{eqnarray}
~ H(\mathcal{A})&=&-(v_++w)\ln(v_++w)-(v_-+w)\ln(v_-+w).  \nonumber
\label{ha:2}
\end{eqnarray}

The joint von Neumann entropy can be derived from the joint pairwise
matrix in (Eq.~(\ref{rho:2})), by which we have
\begin{eqnarray}
(\lambda+y-w)[(w-\lambda)+y][(\lambda-v_{+})(\lambda-v_{-})-|u|^2]=0.
\nonumber
\end{eqnarray}
The eigenvalues ${\lambda_i} (i=1,2,3,4)$ can be obtained analytically.
Then the joint entropy is demonstrated as
\begin{eqnarray}
H(\mathcal{A},\mathcal{B})&=&-(w+y)\ln(w+y)-(w-y)\ln(w-y)  \nonumber \\
&&-\sum_{\lambda=\lambda_\pm}{\lambda}\ln{\lambda}, \nonumber
\end{eqnarray}
where
$\lambda_{\pm}=\frac{1}{2}\{(v_{+}+v_{-}){\pm}[(v_{+}-v_{-})^2+4|u|^2]^{1/2}\}$.

The conditional density ${\rho}_{\mathcal{A}|{\Pi}^{\mathcal{B}}_{k}}$ is
measured by the projections tuned by $\theta$ and $\phi$
\begin{eqnarray}
|\Psi_1{\rangle}_{\mathcal{B}}&=&\cos(\theta)|\downarrow{\rangle}_{\mathcal{B}%
}+e^{i\phi} \sin(\theta)|\uparrow{\rangle}_{\mathcal{B}}~  \nonumber  \label{p:1} \\
|\Psi_2{\rangle}_{\mathcal{B}}&=&e^{-i\phi} \sin(\theta)|\downarrow{\rangle}_{%
\mathcal{B}}-\cos(\theta)|\uparrow{\rangle}_{\mathcal{B}}.  \nonumber  \label{p:2}
\end{eqnarray}
Under such projections, the conditional density matrix is shown as
\begin{eqnarray}
~ {\rho}_{\mathcal{A}|{\Pi}^{\mathcal{B}}_{\Psi_{\alpha}}}&=&|\Psi_\alpha{\rangle}%
_{\mathcal{B}} {\langle}\Psi_\alpha|\{|\downarrow{\rangle}_{\mathcal{A}}{\langle}%
\downarrow|X_{\alpha,+} +|\uparrow{\rangle}_{\mathcal{A}}{\langle}\uparrow|X_{\alpha,-}  \nonumber
 \\
&&+|\downarrow{\rangle}_{\mathcal{A}}{\langle}\uparrow|Y_{\alpha} +|\uparrow{\rangle}_{\mathcal{A}}{%
\langle}\downarrow|Y^{*}_{\alpha}\}/p_{\alpha}. \nonumber
\end{eqnarray}

For ${\alpha}=1$,
\begin{eqnarray}  \label{p:1}
X_{1,+}&=&v_{+}{\cos}^2(\theta)+w{\sin}^2(\theta), \\
X_{1,-}&=&w{\cos}^2(\theta)+v_{-}{\sin}^2(\theta),  \nonumber \\
Y_{1}&=&{\sin}(\theta){\cos}(\theta)[ e^{i\phi}u^{*}+e^{-i\phi}y],  \nonumber
\\
p_{1}&=&w+v_{+}{\cos}^2(\theta)+v_{-}{\sin}^2(\theta).  \nonumber
\end{eqnarray}

For ${\alpha}=2$,
\begin{eqnarray}  \label{p:2}
X_{2,+}&=&v_{+}{\sin}^2(\theta)+w{\cos}^2(\theta), \\
X_{2,-}&=&w{\sin}^2(\theta)+v_{-}{\cos}^2(\theta),  \nonumber \\
Y_{2}&=&-{\sin}(\theta){\cos}(\theta)[ e^{i\phi}u^{*}+e^{-i\phi}y],
\nonumber \\
p_{2}&=&w+v_{+}{\sin}^2(\theta)+v_{-}{\cos}^2(\theta).  \nonumber
\end{eqnarray}
Then the eigenvalues of the conditional density matrix read
\begin{eqnarray}~\label{cdm:1}
~ {\lambda}^{\alpha}_{\pm}(\theta,\phi)&=&\frac{1}{2p_{\alpha}}%
\{(X_{\alpha,+}+X_{\alpha,-}) {\pm}[(X_{\alpha,+}-X_{\alpha,-})^2  \nonumber
\label{p:3} \\
&&+4|Y_{\alpha}|^2]^{1/2}\}.
\end{eqnarray}
The conditional von Neumann entropy is shown as
\begin{eqnarray}
H(\mathcal{A}|\{{\Pi}^{\mathcal{B}}_{k}\})(\theta,\phi)&=&-\sum_{\alpha=1,2} p_{\alpha}[{%
\lambda}^{\alpha}_{+}(\theta,\phi){\ln}{\lambda}^{\alpha}_{+}(\theta,\phi)
\nonumber \\
&&+{\lambda}^{\alpha}_{-}(\theta,\phi){\ln}{\lambda}^{\alpha}_{-}(\theta,%
\phi)].  \nonumber
\end{eqnarray}
Hence
\begin{eqnarray}  \label{p:4}
\delta(\theta,\phi)=H(\mathcal{A})-H(\mathcal{A},\mathcal{B})+H(\mathcal{A}|{%
\Pi}^{\mathcal{B}}). \nonumber
\end{eqnarray}
Finally, the QD can be obtained by optimizing the $\theta$ and
$\phi$ both in the regime $[0,{\pi}/2]$ to minimize
$\delta(\theta,\phi)$, shown as
\begin{eqnarray}
\mathcal{D}=\min_{\{\theta,\phi\}}\{\delta(\theta,\phi)\}. \nonumber
\end{eqnarray}
Consequently, the corresponding classical correlation can also be obtained
by
\begin{eqnarray}
\mathcal{C}=\max_{\{\theta,\phi\}}\{H(\mathcal{A})+H(\mathcal{B})-H(\mathcal{A},\mathcal{B})-\delta(\theta,\phi)\}. \nonumber
\end{eqnarray}

\section{Minimization of the conditional entropy in thermodynamic limit}

In the thermodynamic limit, all non-zero elements of the pairwise density matrix in Eq.~(\ref{rho:2})
have been derived in Eq.~(\ref{parameter:1}).
So we can arrive at Eqs.~(\ref{p:1}) and (\ref {p:2})
\begin{eqnarray}
(X_{k,+}+X_{k,-})&=&{\eta}_{k}(\theta), \nonumber\\
(X_{k,+}-X_{k,-})&=&(2{\beta}^2-1){\eta}_{k}(\theta),  \nonumber \\
|Y_{k}|^2&=&|Y|^2={\beta}^{4}(1-{\beta}^2)^2{\sin}^{2}{2\theta}{\cos}^{2}{%
\phi},  \nonumber \\
p_{k}&=&{\eta}_{k}(\theta),  \nonumber
\end{eqnarray}
where
$\eta_{k}(\theta)=\frac{1}{2}[1+(-1)^{k-1}(2{\beta}^2-1){\cos}{2\theta}]$,
with $k=1,2$.
Therefore the eigenvalues of the conditional density matrix in Eq.~(\ref{cdm:1})
are obtained as
\begin{eqnarray}
{\lambda}^{k}_{\pm}(\theta,\phi)&=&\frac{{\eta}_{k}(\theta){\pm}[(2{\beta}
^2-1)^2{\eta}^2_{k}(\theta)+4|Y|^2]^{1/2}}{2p_{k}} \nonumber \\
&=&\frac{{\eta}_{k}(\theta)}{2p_{k}}\{1{\pm}[(2{\beta}^2-1)^2+\frac{4|Y|^2}{{%
\eta}^2_{k}(\theta)}]^{1/2}\},  \nonumber
\end{eqnarray}
which can be reduced to $
\lambda^{k}_{\pm}(\theta,\phi)=\frac{{\eta}_{k}(\theta)}{2p_k}[1{\pm}
x_k(\theta,\phi)]$  by defining
$x_{k}(\theta,\phi)=[(2{\beta}^2-1)^2+\frac{4|Y|^2}{{\eta}
^2_{k}(\theta)}]^{1/2}$ $(0{<}x_{k}{<}1)$. The conditional von
Neumann entropy is described as
\begin{eqnarray}
&&H(\mathcal{A}|\{\Pi^{\mathcal{B}}\})(\theta,\phi)  \nonumber \\
=&&\sum_{k=1,2}\{{\eta_k}(\theta)\ln{\eta_k}(\theta)  \nonumber \\
&&-[\frac{{\eta}_{k}(\theta)}{2}(1{+}x_k(\theta,\phi))]{\ln} [\frac{{\eta}%
_{k}(\theta)}{2}(1{+}x_k(\theta,\phi))]  \nonumber \\
&&-[\frac{{\eta}_{k}(\theta)}{2}(1{-}x_k(\theta,\phi))]{\ln} [\frac{{\eta}%
_{k}(\theta)}{2}(1{-}x_k(\theta,\phi))]\}  \nonumber \\
=&&{\ln}{2}-\sum_{k=1,2}\frac{\eta_k(\theta)}{2} [(1{+}x_k(\theta,\phi)){%
\ln}(1{+}x_k(\theta,\phi))  \nonumber \\
&&+(1{-}x_k(\theta,\phi)){\ln}(1{-}x_k(\theta,\phi))],\nonumber
\end{eqnarray}
Since $F(x)=-[(1+x)\ln(1+x)+(1-x)\ln(1-x)]$ is the monotonically
decreasing function, $\phi$ is selected to $0$ to maximize $
x_{k}(\theta,\phi)$. Furthermore, we find $\theta=\pi/4$ to minimize
the conditional von Neumann entropy as~\cite{luo1}
\begin{eqnarray}
H(\mathcal{A}|\Pi^{\mathcal{B}})&=&\ln(2)-\frac{1}{2}[(1+M)\ln(1+M)~
\nonumber \\
&&+(1-M)\ln(1-M)],\nonumber
\end{eqnarray}
where
$M=x(\pi/4,0)=[(2{\beta}^2-1)^2+16{\beta}^4(1-{\beta}^2)^2]^{1/2}$.

\end{document}